\documentclass[12pt,preprint]{aastex}
\usepackage{amsmath}
\usepackage{amsbsy}
\usepackage{rotating}

\slugcomment{Draft: {\today}}

\shorttitle{Composition Structure of ICMEs}
\shortauthors{Reinard et al.}

\begin{document}

%% LaTeX will automatically break titles if they run longer than
%% one line. However, you may use \\ to force a line break if
%% you desire.

\title{Composition Structure of Interplanetary Coronal Mass Ejections From Multispacecraft Observations, Modeling, and Comparison with Numerical Simulations}

%% Use \author, \affil, and the \and command to format
%% author and affiliation information.
%% Note that \email has replaced the old \authoremail command
%% from AASTeX v4.0. You can use \email to mark an email address
%% anywhere in the paper, not just in the front matter.
%% As in the title, use \\ to force line breaks.

\author{Alysha A. Reinard\altaffilmark{1},
        Benjamin J. Lynch\altaffilmark{2}, 
    and Tamitha Mulligan\altaffilmark{3}}

%\author{R. J. Hanisch\altaffilmark{5}}
%\affil{Space Telescope Science Institute, Baltimore, MD 21218}

%% Notice that each of these authors has alternate affiliations, which
%% are identified by the \altaffilmark after each name.  Specify alternate
%% affiliation information with \altaffiltext, with one command per each
%% affiliation.
%
\altaffiltext{1}{University of Colorado/Cooperative Institute for Research in Environmental Sciences and National Oceanic and Atmospheric Administration/Space Weather Prediction Center, Boulder, CO, USA, Boulder, CO 80305, USA; alysha.reinard@noaa.gov}
\altaffiltext{2}{Space Sciences Laboratory, University of California,
        Berkeley, CA 94720, USA; blynch@ssl.berkeley.edu}
\altaffiltext{3}{Space Sciences Department, The Aerospace Corporation, Los Angeles, CA 90009, USA; tamitha.mulligan@aero.org}

\begin{abstract}

We present an analysis of the ionic composition of iron for two interplanetary coronal mass ejections observed in May 21-23 2007 by the ACE and STEREO spacecraft in the context of the magnetic structure of the ejecta flux rope, sheath region, and surrounding solar wind flow. This analysis is made possible due to recent advances in multispacecraft data interpolation, reconstruction, and visualization as well as results from recent modeling of ionic charge states in MHD simulations of magnetic breakout and flux cancellation CME initiation.  We use these advances to interpret specific features of the ICME plasma composition resulting from the magnetic topology and evolution of the CME. We find that in both the data and our MHD simulations, the flux ropes centers are relatively cool, while charge state enhancements surround and trail the flux ropes. 
The magnetic orientation of the ICMEs are suggestive of magnetic breakout-like reconnection during the eruption process, which could explain the spatial location of the observed iron enhancements just outside the traditional flux rope magnetic signatures and between the two ICMEs. 
Detailed comparisons between the simulations and data were more complicated, but a sharp increase in high iron charge states in the ACE and STEREO-A data during the second flux rope corresponds well to similar features in the flux cancellation results.  We discuss the prospects of this integrated in-situ data analysis and modeling approach to advancing our understanding of the unified CME-to-ICME evolution.

\end{abstract}

\keywords{Sun: solar-terrestrial relations --- Sun: coronal mass ejections (CMEs) --- 
          Sun: flares --- Sun: heliosphere --- Sun: solar wind}

%%%%%%%%%%%%%%%%%%%%%%%%%%%%%%%%%%%%%%%%%%%%%%%%%%%%%%
%
%
\section{Introduction}

Coronal mass ejections (CMEs) are periods of explosive magnetic energy release and coronal field reconfiguration that blow out huge portions of the quasi-stable solar atmosphere into interplanetary space \citep{Tousey1973,Gosling1997}. When these solar eruptions are Earth-directed, they generally arrive in 1--5 days \citep{Cane2003} and their impact can cause geoeffective space weather responses \citep{Gosling1993, Zhang2004}. CMEs are observed soon after their eruption by remote sensing instruments that capture the global structure of these events. Their morphological structure in coronagraph observations typically consists of a bright bubble of plasma, sometimes with a darker cavity and bright core \citep[e.g.,][]{Illing1986,Howard1997}. Once these CMEs propagate into the heliosphere they are commonly referred to as interplanetary CMEs (ICMEs) and are observed in-situ by spacecraft that sample local measurements of the magnetic field, plasma, and composition. ICME identification is based on a variety of parameters including enhancements in magnetic field,  composition and temperature depressions \citep[e.g., see][and references therein]{Zurbuchen2006}. 

Though CMEs and ICMEs are the same phenomena, the remote sensing and in-situ observations can often be difficult to directly relate to one another, due to both the nature of the observations as well as the evolution that takes place during the transit from the Sun to the spacecraft. In-situ solar wind composition data provides an important connection between the ICME observations and associated observations at the Sun, in particular providing a measure of physical properties in the corona that can be compared to EUV and X-ray emission and white light coronagraph observations. Charge state composition is directly related to the electron temperature at the source region, with hotter source material resulting in higher in-situ charge states \citep{Buergi1986}. During the solar wind expansion, these charge states Òfreeze-inÓ once the expansion timescale becomes larger than the collision timescale \citep{Hundhausen1968}. 
{ICMEs often contain unusual charge state distributions. These charge states can be enhanced relative to the ambient solar wind due to heating at the source region, lower than the ambient solar wind due to incompletely heated cold prominence material, or a combination of both low and high charge states \citep{Bame1979, Gloeckler1999, Richardson2004,  Lepri2010, Gruesbeck2011, Gilbert2012}. Because of this complexity, it is important to examine the full charge state distribution.
Charge state enhancements are weakly correlated with associated flare magnitude \citep{Reinard2005} and tend to be strongest near the center of the ICME \citep{Reinard2008}. In this paper we examine the in-situ iron charge state data  from the ACE and STEREO spacecraft and compare them with theoretical predictions derived from MHD simulations of two different CME initiation models using the technique developed by \citet{Lynch2011}. 

Given the localized nature of ICME observations (ICMEs are typically observed by a single spacecraft that samples a narrow trajectory through the larger CME structure) drawing conclusions about the global structure of ICMEs can be complicated as it is not always clear which part of the ICME is being sampled. The subset of ICMEs with the most structured field and plasma signatures are known as magnetic clouds and are defined as having an enhanced field strength, relatively smooth magnetic field rotations, and a lower than expected temperature \citep{Burlaga1981}. While the commonly accepted occurrence rate for magnetic clouds is between 30--50\% of all ICME events \citep{Gosling1990,Richardson2004}, there is increasing evidence that all or most CMEs develop or contain some sort of flux rope structure \citep{Jian2006,Kilpua2011}. 
For those ICMEs, or portions of ICMEs, that meet the criteria to be considered magnetic clouds \citep{Burlaga1981} the overall ICME structure can be approximated by modeling the in-situ ICME magnetic fields as force-free, cylindrical flux ropes \citep[e.g.,][]{Lepping1990}. These techniques allow researchers to interpret the cavity portion of the CME coronagraph morphology with the in-situ magnetic flux rope structure \citep[][]{Illing1986,Wood1999,Webb2009,Vourlidas2012} and indeed observations have confirmed this association \citep[e.g.][]{DeForest2011}. However, analysis of the current density structure, measures of the interplanetary pressure-balance, and multispacecraft observations have shown that even magnetic clouds are not precisely force-free, cylindrical objects. Therefore non-force free and consequently non-cylindrical fits were developed \citep[e.g.,][]{Farrugia1993,Farrugia1999,Mulligan2001,Hidalgo2002} that allowed more complexity in the local ICME field modeling, while still retaining a big picture view of the global structure. Though these advances have allowed a greater number of ICMEs to be modeled, all of these analytic flux rope modeling techniques are limited in that they are only relevant for the flux rope-like portion of the ICME; non-flux rope ICMEs or cuts through an ICME that do not cleanly intersect the flux rope cannot be included because they lack the proper magnetic structure needed for the fits. In theory, the pressure-balance Grad-Shafranov reconstruction techniques \citep[e.g.,][]{Hu2001,Hu2002,Isavnin2011} avoid the a-priori flux rope criteria, as well as the problematic identification of the flux rope boundaries, but in practice are most successful when reconstructing flux rope structures \citep{Riley2004,Al-Haddad2011}.  Recently, \citet{Mulligan2012} addressed this issue by developing a technique to incorporate non-MC (magnetic cloud) material into MC flux rope modeling.

Multispacecraft observations of ICMEs have made and continue to make a valuable contribution to our understanding of the larger heliospheric structure, placing important constraints on flux rope modeling and the spatial extent of ICME ejecta over which the coherent flux rope signatures persist \citep{Burlaga1981,Hammond1995,Mulligan1999,Mulligan2001,Riley2003}. The Solar-Terrestrial Observatory (STEREO) mission \citep{Kaiser2008} was launched in October 2006 and consists of twin spacecraft at 1 AU traveling ahead of and behind the Earth at a rate of approximately 22.5 degrees per year. The STEREO observing geometry provides a means of relating multispacecraft in-situ observations to each other and to the continuous remote coronal and heliospheric imaging starting at the Sun and extending to 1 AU. The data from the SECCHI suite \citep{Howard2008} combined with the IMPACT \citep{Luhmann2008} and PLASTIC \citep{Galvin2008} in-situ field and plasma measurements have ushered in a new era of direct CME-ICME observations, explicitly linking various morphological features to their associated in-situ properties \citep[e.g.,][]{Harrison2008,Harrison2010,Davis2009,Davies2009,Moestl2009a,Roulliard2009,Webb2009,Lynch2010,DeForest2011}. In addition, the two STEREO spacecraft, along with the ACE and WIND spacecraft, offer three or four tracks through a given ICME rather than just one, allowing more of the ICME spatial structure to be sampled \citep{Kilpua2011}. However, often even three tracks of observations through an ICME cannot fully constrain the ICME global structure and further heliospheric modeling is necessary to interpret and relate the multispacecraft data back to remote observations at the Sun. \citet{Mulligan2012} describe a sophisticated method that provides an interpolated spatial mapping of the data between the spacecraft tracks. This method provides an estimate of the global ICME structure and simplifies the comparison with modeled and observed CME structures near the Sun.  

The structure of the paper is as follows. In Section 2 we describe our methodology. First, in section 2.1, we briefly review previous work on the source region and the multispacecraft in-situ data of the 2007 May 21--23 ICMEs. In section 2.2 we describe the \citet{Mulligan2012} reconstruction and interpolation techniques for transforming the in-situ timeseries data into spatial maps and present two-dimensional composition data in the context of the ICME flux rope structures and their surrounding regions. In section 2.3 we describe the \citet{Lynch2011} application of ionic charge state calculations to numerical MHD simulations to derive theoretical spatial distributions of iron charge states for two idealized CME initiation scenarios. In section 3 we compare the observed and simulated ionic composition structure; first, on the global scale of the two flux rope ICMEs and their  spatial structure, and second, by constructing a set of synthetic spacecraft trajectories through the MHD simulation data for direct comparison with the multispacecraft measurements of the detailed iron charge state distributions (Fe$^{+6}$ to Fe$^{+20}$). In section 4 we discuss how the structure and complexity of the interplanetary composition data is reflected in the complexity of the May 19 and 20 CME source region's magnetic topology and resulting eruption scenario.  We conclude, in section 5, with a summary of our results and prospects for future analyses.

\section{Methodology, Data Analysis, and Numerical Modeling}

\subsection{Event Description and Multispacecraft Observations}

The 2007 May 21--23 ICMEs occurred when STEREO-A and STEREO-B were approximately 9 degrees apart with a separation from ACE of 5.9 and 3.1 degrees, respectively. These events both originated in AR10956, which was located nearly exactly at disk center. The first event was associated with a B9.5 flare that peaked at 13:02 UT on May 19, while the second event was associated with a B6.7 flare that peaked at 05:56UT on May 20.  The first of these events is well studied. \citet{Li2008} examined the source region topology and the photospheric magnetic field evolution leading up to the eruption. \citet{Bone2009} examined the pre-eruption filament formation and \citet{Liewer2009} analyzed the stereoscopic 3-dimensional trajectory during the filament lift-off and eruption. \citet{Veronig2008}, \citet{Gopalswamy2009} and \citet{Attrill2010} characterized the EUV-dimming and coronal wave signatures. \citet{Kerdraon2010} examined the CME-ICME connection in terms of the local acceleration of energetic electrons during the eruption and their interplanetary radio burst signatures. 
The left column of Figure~\ref{fig1} shows the 171\AA\ EUV dimming associated with the May 19 eruption.  In the top panel is the base image, taken at 12:22UT, in which the active region is the bright region seen in the center.  In the bottom panel is the difference image \citep[13:12UT-12:22UT, technique described in][] {ReinardB2008} with the dark dimming region indicating the erupting material from the northwest section of the active region \citep[see also][]{Li2008,Kilpua2009}.  The right column in Figure~\ref{fig1} shows similar images for the May 20 eruption (04:52UT base image, top panel, and 06:02UT-04:52UT difference image, bottom panel).  In this case, the eruption originated from the southeast of the same active region. \citet{Mierla2008} utilized STEREO/COR1 height-time diagrams to analyze the propagation direction of the May 20 eruption.

\citet{Kilpua2009} discussed the CME-ICME connection during this period associating the CDAW catalog entry of a fast CME occurring at 13:24 UT (speed of 958 km/s) with the well-structured magnetic cloud observed by STEREO, ACE, and WIND on May 22. A second, slower CME was observed at 13:48UT (speed of 294 km/s) and associated by \citet{Kilpua2009} with a second magnetic cloud seen by STEREO-A on May 23. Following  \citet{Kilpua2009} we denote these events as MC1 and MC2.
Figure~\ref{fig2} shows the STEREO-B, ACE, and STEREO-A in-situ data (from left to right) for the 2007 May 21--23 period (DOY 141--143), which includes ICME signatures resulting from both eruptions. From top to bottom, we plot the bulk radial proton velocity $v_p$, proton number density $n_p$, proton temperature $T_p$ (black line) and the expected temperature $T_{\rm exp}$ \citep[green line;][]{Lopez1987}, the distribution of iron charge states (Fe$^{+6}$ to Fe$^{+20}$, denoted as $Q_{Fe}$), and the interplanetary magnetic field in RTN coordinates: $B_{\rm R}$ (red), $B_{\rm T}$ (green), $B_{\rm N}$ (blue), and $B$ magnitude (black). In each panel the MC1 interval is indicated by the yellow vertical lines.  Spacecraft data that clearly transverse the magnetic flux rope portion of the ICME are shaded. The MC2 interval is likewise indicated in light blue lines and shading. 

Both \citet{Liu2008} and \citet{Kilpua2009} performed Grad-Shafranov fits for MC1 obtaining moderately inclined flux rope orientations with reasonably circular cross-sections. \citet{Moestl2009b} obtained similar results applying a multispacecraft Grad-Shafranov fit to STEREO-B and WIND data, while \citet{Moestl2009c} fit MC2 using Grad-Shafranov modeling at STEREO-A, which was then compared and optimized using WIND observations to yield a distorted, slightly flattened, ellipse.

\citet{Rakowski2011} examined the ionic charge states during MC1 and modeled the evolution of the flux rope ejecta with an analytic spheromak magnetic field configuration. They calculated a predicted ionic charge state distribution resulting from internal ICME heating supplied via the dissipation of magnetic energy associated with the ICME magnetic structure relaxing toward a force-free equilibrium. This approach was able to produce Fe$^{+8}$ to Fe$^{+13}$ as well as other moderate charges states in C, O, and Si, but not the observed shapes of the full charge state distributions. We describe the \citeauthor{Rakowski2007} (\citeyear{Rakowski2007}, \citeyear{Rakowski2011}) procedure for the calculation of ionic charge states and our application of it to our MHD simulations of CME initiation and flux rope formation in Section 2.3.

\subsection{Spatial Mapping and In-Situ Data Interpolation}

\citet{Mulligan2012} have recently developed important extensions to multispacecraft data analysis techniques. First, by extending the \citet{Mulligan2001} non-cylindrical flux rope model to include the bulk plasma velocity vector into the model inversion and fitting, and second, by using the velocity time series in multispacecraft data to create interpolated spatial maps of the solar wind and ICME plasma structures. 
By incorporating the velocity flow deflections around the ICME obstacle into the calculation of the orientation and size of the flux rope portion of the ICME, we can utilize multispacecraft data clearly associated with the envelope of the ICME but not the flux rope itself in both the ICME modeling and the interpretation of the surrounding heliospheric solar wind structure. The spatial mapping procedure consists of first integrating the components of the bulk velocity time series in each of the multispacecraft data tracks to obtain the relative spatial positions of any measured or calculated in-situ quantities of interest and then employing Delauney triangulation to interpolate between each of these spacecraft tracks. The \citet{Mulligan2012} spatial mapping procedure is derived from interpolation techniques that have been used in the analysis of Cluster data \citep{Chanteur1993,Chanteur1998,Sibson1981} and represents a substantial improvement over the simplified version of the spatial mapping used in \citet{Lynch2003} and \citet{Reinard2010}.
\citet{Mulligan2012} constructed spatial contour maps from the STEREO, ACE, and WIND time series data during the 2007 May 21--24 period to analyze the ICME and surrounding solar wind structure of MC1 and MC2 in the bulk plasma velocity, proton density and temperature and He$^{++}$/H$^+$ ratio (see their Figure~9).
 
In Figure~\ref{fig3} we plot the spatial mapping of the ACE/SWICS and STEREO/PLASTIC ionic composition measurements of the average iron charge state $\langle Q_{\rm Fe} \rangle$. In this plot the time series data is converted to spatial coordinates with the origin of the spatial map located at the Earth's position in the y direction and centered between the two flux ropes in the x direction \citep[as in][]{Mulligan2012}.  The axes are given in AU and the individual spacecraft tracks are shown in grey, with STEREO-B (STB) tracing the bottom edge of the spatial map, ACE in the center, and STEREO-A (STA) tracing the top edge. Each spacecraft's position in the plot increases from right to left in time and the Sun is to the left.  
The black arrow vector field regions are the GSE (Geocentric Solar Ecliptic) $B_x$ and $B_y$ components of the interpolated magnetic field showing the large-scale rotation associated with the cross-sections of the MC1 and MC2 flux rope structures. Specifically, the vector arrows drawn along the spacecraft track are the observed data in the yellow (blue) intervals shown in Figure~\ref{fig2} for MC1 (MC2) and the arrows between spacecraft tracks are determined by the \citet{Mulligan2012} interpolation method.  
From the spatial mapping of the flux rope structure it is clear that STB passes near the center of MC1 and is either at or just outside the boundary of the MC2 flux rope; ACE passes comfortably through both flux ropes, but has a slightly higher impact intersection at MC1 with respect to the ``center" of the cross-sectional field rotation; STA just misses the edge of MC1 and has a relatively high-impact angle intersection with MC2. As discussed by \citet{Mulligan2012}, the flux rope structures inferred by the spatial mapping interpolation are completely consistent with various multispacecraft flux rope modeling results obtained by other researchers \citep{Liu2008,Kilpua2009,Moestl2009b,Moestl2009c}.  

Examining Figure~\ref{fig3} we see low values of $\langle Q_{\rm Fe} \rangle$ in the front (right) portion of MC1 and at the flank that crosses STA. There is an increase in $\langle Q_{\rm Fe} \rangle$ in the back half of MC1 along the  ACE trace through the data. Interestingly, STB, which passes most closely through the center of the flux rope, does not detect any $\langle Q_{\rm Fe} \rangle$ enhancements at all.  For MC2 the center of the flux rope has $\langle Q_{\rm Fe} \rangle$ values that are significantly lower than the rest of the flux rope. The highest values of $\langle Q_{\rm Fe} \rangle$ in MC2 occur at the STA flank and following the event.  Also of interest is the $\langle Q_{\rm Fe} \rangle$ enhancement that seems to link the two events, extending from the back half of MC1, filling the space in between the two events, and blending into the enhancement at the STA flank of MC2.

We have used $\langle Q_{\rm Fe} \rangle$ in the spatial mapping because
it is a good proxy for iron charge state enhancements and indicative of
ICME material \citep[e.g.,][]{Lepri2004}, however there are limitations to
using just the distribution average alone. By comparing the timeseries of
the $Q_{\rm Fe}$ distributions shown in Figure 2 with the Figure 3 spatial
map, we see that there are two different signatures that give rise to
similar enhancements in $\langle Q_{\rm Fe} \rangle$. For example, the ACE
MC1 flux rope and STA data preceding the MC2 flux rope have a $\langle
Q_{\rm Fe} \rangle$ enhancement due to a general broadening of the entire
distribution to moderately higher charge states, whereas the STA MC2
$\langle Q_{\rm Fe} \rangle$ enhancement arises because of a bimodal
$Q_{\rm Fe}$ signature that includes a significant peak at the high iron
charge states of +15/+16.

\subsection{Inferring ICME Charge States from MHD Modeling}

\citet{Lynch2011} demonstrated a post-processing procedure in which the ionic charge state composition of heavy ions is calculated from MHD simulations of CME initiation using the single-fluid (proton) velocity, temperature and density evolution. Starting with a snapshot late in the modeling run,  a set of individual fluid parcels are integrated backwards in time using the MHD velocity field to derive the temperature and density history of these parcels. The MHD temperature and density profiles are then used as inputs into the \textit{Blast Propagation in a Highly Emitting Environment} code \citep[\textit{BLASPHEMER};][]{Laming2002,Laming2003,Rakowski2007} which solves the continuity equations for heavy ion charge states of interest,
\begin{equation}
\frac{\partial n_q}{\partial t} = n_e \left[ C^{\rm ion}_{q-1} n_{q-1} - \left( C^{\rm ion}_q + C^{\rm rr}_q + C^{\rm dr}_{q}\right) n_q + \left( C^{\rm rr}_{q+1} + C^{\rm dr}_{q+1} \right) n_{q+1} \right] \; ,
\end{equation}
where $n_e$ is electron density and $n_q$ is the density of a given charge state. 
Here, for a charge state $q$, the right hand side represents the sources and sinks of this charge state from electron impact ionization ($C^{\rm ion}$), and radiative ($C^{\rm rr}$) and dielectronic ($C^{\rm dr}$) recombination. This approach makes the following assumptions: (1) the heavy ions act as passive tracers of the MHD flow, (2), the plasma remains electrically neutral $n_e=n_p$ (where $p$ refers to protons), and (3), the electron temperature is given by $T_e=T_p$. 
\citet{Lynch2011} applied this technique to two different axisymmetric MHD simulations run with different codes; a magnetic breakout CME run with the \textit{ARC7} code \citep{Allred2008} and a flux cancellation CME run with the \textit{Magnetohydrodynamics-on-A-Sphere} (\textit{MAS}) code \citep{Linker2001,Lionello2009}. The comparison of the derived charge states of the \textit{ARC7} magnetic breakout simulation and the \textit{MAS} flux cancellation simulation provided strong evidence that the eruptive flare heating in the low corona could produce enhancements in commonly measured high charge state ratios (e.g., O$^{7+}$/O$^{6+}$, Fe$^{\ge16+}$/Fe$_{\rm total}$) within the CME flux rope that would be carried out into the heliosphere. 
As mentioned in \citet{Lynch2011}, these idealized simulations were not designed for the ionic charge state analysis we performed or tailored to model specific CME events, so the background solar wind solutions and resulting CME properties are not identical. Thus, the simulation and observational comparisons in this paper are, for the most part, qualitative.
The charge state analyses in these simulations provide us with the full iron charge state distribution, so we can construct both the range of the distribution available in the in-situ data (Fe$^{+6}$ to Fe$^{+20}$) as well as the average iron charge states $\langle Q_{\rm Fe} \rangle$ for comparison to the spatial mapping in Figure~\ref{fig3}.  
In Figure~\ref{fig4} we plot the average iron charge state $\langle Q_{\rm Fe} \rangle$ derived from the magnetic breakout model simulation (left panel; hereafter BM) and the flux cancellation model simulation (right panel; hereafter FC). The $\langle Q_{\rm Fe} \rangle$ contour maps are each made up of a 60-by-60 grid of Lagrangian fluid parcels where we have constructed $\langle Q_{\rm Fe}\rangle =\sum{ \left( qn_q/\sum{ n_q } \right) }$ from the full distribution of iron charge states in each pixel. Representative flux contours are over-plotted showing the magnetic field cross sections of the flux rope ejecta, their sheath regions, and the surrounding solar wind structure. We note that the BM simulation results herein are presented with a flipped $y$-axis \citep[compared to][]{Lynch2011} to ease the visual association with the spatial mapping results.

\section{Comparison of Observations and Simulation Results}

\subsection{Large-Scale $\langle Q_{\rm Fe} \rangle$ Structure}

The large-scale heliospheric structure of the average iron charge state signatures associated with the May 21--23 successive ICMEs in Figure~\ref{fig3} have characteristics found in both the BM and FC MHD simulation results.  
MC1 has only slightly enhanced average iron charge states of $\langle Q_{\rm Fe} \rangle \sim$10.5 in the ACE measurements in the back half of the flux rope. 
MC2 has stronger enhancements, up to $\langle Q_{\rm Fe} \rangle$ $\sim$12, that are seen in both the STA and ACE trajectories through the flux rope. In both MC1 and MC2 the location of the average iron enhancements are in the back half of the actual flux ropes, offset from their respective symmetry axes (flux rope center), and also in the trailing solar wind flow following the rear boundaries of the flux rope field rotations. 
Looking at Figure~\ref{fig4}, both the BM and FC simulations have a similar qualitative structure in terms of enhancements in the latter half of the flux rope resulting from the contribution of flare-heated plasma being supplied by the eruptive flare reconnection jet. The magnitude of the charge state enhancements in the FC simulation are much larger, including $\langle Q_{\rm Fe} \rangle$ values of $\gtrsim$15 in the reconnection jet outflow and result in a shell of $\langle Q_{\rm Fe} \rangle \sim$12--13 enhancement encircling the flux rope's lower charge state core.
In the BM simulation results there is also a clear enhancement preceding the flux rope in the topologically well defined sheath region due primarily to compressional heating of the density enhancements and in some part to the breakout reconnection heating during the eruption.  This compares well with the charge state enhancement ahead of MC2 which extends well beyond the flux rope boundaries.

We can also look at Figure~\ref{fig3} in the context of the source region EUV dimmings of Figure~\ref{fig1}. For MC1, the charge state enhancements are minimal and follow the ICME flux rope. We see that in the MC1 source region, the flare occurs on the eastern side of the eruption (dimming) which would have crossed 1AU beyond the STB spacecraft and so is not seen in the in-situ observations (i.e., the half of the MC1 flux rope not included in the $\langle Q_{\rm Fe} \rangle$ spatial map).  
For MC2 the bulk of the enhancements are in the western half of the event, which corresponds well to location of the flare in relation to the CME source region.  The dimming associated with MC2 occurs rather strongly to the western side of the active region in which the flare occurs.  The correspondence of the flare location and EUV dimming offset is consistent with the idea that the flare is indeed responsible for the CME heating that results in the enhanced charge states.

\subsection{Multispacecraft and Simulation Magnetic Field Structure}

Inspired by the general agreement of the large-scale average iron charge state signatures in the multispacecraft spatial mapping and the results from the MHD CME modeling, we now extend our comparison to a broader range of the iron charge state distributions sampled along synthetic spacecraft trajectories through the simulation results. The MC1 and MC2 flux ropes are each intersected by two of the three spacecraft and have the third spacecraft track either right at or just outside the flux rope boundary. The multispacecraft in-situ data yield six ICME intervals to be examined. In the top panel of Figure~\ref{fig5} \citep[adapted from][]{Mulligan2012} we plot the STEREO and ACE spacecraft trajectories through the MC1 and MC2 flux rope ecliptic plane/spatial map field rotations. The MC1 (MC2) ellipse colored yellow (blue) show the boundaries of the \citet{Mulligan2012} multispacecraft elliptical flux rope model fits to the ICME flux rope interval. Figure~\ref{fig5} lower-left and lower-right panels show the BM and FC simulation snapshots of representative flux surfaces and the set of synthetic spacecraft trajectories that roughly correspond to the relative position and angle of the STB, ACE, and STA tracks through MC1 and MC2. 
For MC1 (MC2) we chose the synthetic spacecraft trajectories at an angle of $+22^{\circ}$ ($-12^{\circ}$) with respect to the simulation's circular flux rope cross section to qualitatively match the multispacecraft observing geometry.
Thus, for each MHD simulation we define six model cuts to compare with the MC1 and MC2 data. 
For example, for the MC1 interval we have the MC1$_{\rm STB}$, MC1$_{\rm ACE}$, and MC1$_{\rm STA}$ in-situ data to be compared with the breakout model trajectories $\{$BM$^{(1)}_{\rm STB}$, BM$^{(1)}_{\rm ACE}$, BM$^{(1)}_{\rm STA}$$\}$ and the flux cancellation trajectories $\{$FC$^{(1)}_{\rm STB}$, FC$^{(1)}_{\rm ACE}$, FC$^{(1)}_{\rm STA}$$\}$.

Since the simulation trajectories represent spatial sampling through the BM and FC results and we are using the in-situ data time series, we want to ensure that the position of the simulation cut and time of the in-situ data correspond, as much as possible, to the same relative position with respect the magnetic flux ropes and the surrounding ICMEs.  
MC1 and MC2 have different flux rope orientations \citep[as seen in Figures~\ref{fig2} and \ref{fig5} of this paper and Figures 8 and 9 of][]{Mulligan2012}, including axial field directions of opposite sign in the two events (seen largely as the sign of the spacecraft $B_N$ component). Therefore, we have transformed the MHD simulation magnetic field data so that the sense of the field rotations and the direction of the axial component in the simulations match each of the MC1 and MC2 events. In the axisymmetric MHD simulations, the ($B_r$, $B_\theta$) field contain the azimuthal (twist) component of the flux rope ejecta while the $B_\phi$ component represents the axial flux rope field. Multiplying the $B_\phi$ component by $-1$ changes the orientation of the axial field but also changes the handedness of the flux rope unless the ($B_r$, $B_\theta$) components are also multiplied by $-1$.

Figures~\ref{fig6a} and \ref{fig6b} plot the entire set of MC1 and MC2  multispacecraft in-situ magnetic field and iron charge state data and the corresponding BM and FC simulation trajectories. The upper panels show the in-situ magnetic field components in each ICME interval, colored in the same manner as Figure~\ref{fig2}, i.e.  $B_{\rm R}$ (red), $B_{\rm T}$ (green), $B_{\rm N}$ (blue), $B$ (black), and the fields from the MHD simulation trajectory sampling $B_r$ (red), $B_\theta$ (green), $B_\phi$ (blue), $B$ (black). Obviously the simulation field magnitudes at $\sim$12$R_{\odot}$ and the in-situ 1AU field magnitudes are very different. 
We have selected plot ranges to allow a quick visual comparison to assess the simulation flux rope signatures with respect to the in-situ magnetic field. Here the spacecraft data are all plotted on the same $\pm$20~nT range whereas the plot ranges for the fields along the BM simulation trajectories are $\pm0.02$~G and along the FC simulation trajectories are $\pm0.08$~G. 
Once again, and by construction, the agreement is quite reasonable. Both the BM and FC simulations capture the qualitative features of the flux rope rotation, at least in the two sampling trajectories that intersect the flux rope in each event: MC1$_{\rm STB}$, MC1$_{\rm ACE}$ and MC2$_{\rm ACE}$, MC2$_{\rm STA}$. This gives us confidence that the comparison of the simulation-derived ionic composition signatures and the in-situ ionic composition signatures is reasonable and can be used to both interpret the observations and evaluate the simulations.

\subsection{Multispacecraft and Simulation $Q_{\rm Fe}$ Distributions}

%============================
% MC1 
%============================
The lower panels in Figures~\ref{fig6a} and \ref{fig6b} plot the detailed iron charge state distributions of the in-situ data and from the MHD simulation trajectory sampling. For the Figure~\ref{fig6a} MC1 comparison, the ordering from top to bottom of STB, ACE, and STA represent cuts through the center, mid-range, and just missing the boundary of the magnetic flux rope. 
The in-situ $Q_{\rm Fe}$ distributions are consistently low throughout MC1$_{\rm STB}$, ranging primarily from +8 to +9. For MC1$_{\rm ACE}$  the charge states have a decreasing profile starting with charge states of +9 to +12 and ending with charge states of mainly +8 and +9.  Charge states in MC1$_{\rm STA}$ range from +8 to +10 at the beginning and end of the event and drop slightly to +7 to +9 at the center of the interval. 

Here, the BM simulation shows the general trend of the center of the flux rope having lower iron charge states: the peak of the distribution dips most strongly in BM$^{(1)}_{\rm STB}$, more moderately in BM$^{(1)}_{\rm ACE}$, and exhibit no variation in BM$^{(1)}_{\rm STA}$. The background solar wind BM charge state distribution outside of the flux rope and sheath region is strongly peaked at +11, which is slightly higher than the no-enhancement levels seen during the ICME periods. 
The FC simulation charge states are more varied reflecting the intensity of the shell of flare heated material and corresponding charge state enhancement in the back of the flux rope structure. While the no-enhancement level of the FC distribution (+7 to +8) tends to be lower than the in-situ observations, for the simulation trajectories FC$^{(1)}_{\rm STB}$ and FC$^{(1)}_{\rm ACE}$ we see a sharp transition to high iron charge states of $+16$ and a more gradual fall off of the level of enhancement in the trailing half. 
The ring structure in the enhancements is also visible in the FC cuts, with FC$^{(1)}_{\rm ACE}$ showing a narrow spike of +16 enhancement in the first half of the flux rope, but also as a weak signal in the FC$^{(1)}_{\rm STB}$ and FC$^{(1)}_{\rm STA}$ plots.
In the FC$^{(1)}_{\rm STA}$ trajectory there is also a remnant of the trailing half enhancement with the distribution peak shifting slightly higher to +8 to +9 and broadening to include trace amounts of +10 to +11.

%============================
% MC2 
%============================
For the Figure~\ref{fig6b} MC2 comparison, STB just misses the flux rope, while ACE and STA represent cuts through the center and mid-range impact angles.  
The in-situ $Q_{\rm Fe}$ distributions in MC2 have more structure and much higher levels of enhanced charge states. While MC2$_{\rm STB}$ is still low (+8 to +9 peak), the distribution is broader with trace amounts of +11 to +13. The latter half of the MC2$_{\rm ACE}$ distributions include a moderate amount of +11 to +13 and a small peak at +16. MC2$_{\rm STA}$ exhibits the most elevated iron charge states with a bimodal structure including a strong peak at +15 to +16 and +8 to +9 at the center of the flux rope interval lasting $\sim$8~hours. The bimodal iron charge state structure is common feature of ICME events with flare heated composition enhancements \citep{Gruesbeck2011, Gilbert2012}.

The BM simulation charge state data for the MC2 comparisons show the same trend as the MC1 cuts in terms of distance from the flux rope center; BM$^{(2)}_{\rm ACE}$ shows the largest central lower charge state dip, BM$^{(2)}_{\rm STA}$ shows a shallower dip, and BM$^{(2)}_{\rm STB}$ shows very little variation. Again, the baseline BM distribution peaks are slightly higher than the non-enhanced periods of the MC2 distributions.
The iron charge state distributions in the FC simulation match features of the MC2 in-situ data reasonably well for the two flux rope intersections. Here the MC2$_{\rm ACE}$ enhancements and the FC$^{(2)}_{\rm ACE}$ distributions appear in the same trailing region of the flux ropes and the gradual return of the iron charge state distribution from +16 to lower enhancement levels are similar. The FC$^{(2)}_{\rm STA}$ sampling trajectory stays in the flare heated enhancement ring longer and therefore shows the most extended hot iron +16 peak in the center of the magnetic field rotation signatures. The relative duration of this central enhancement is qualitatively similar to the MC2$_{\rm STA}$ data. The FC$^{(2)}_{\rm STA}$ enhancement region also has a bimodal distribution, with the low charge state peak at +8 in the first half of the region then elevating to higher charge states about one third of the way through the event and staying elevated until about two thirds of the way through the event.  This structure is qualitatively similar to what is seen in the STA observations.

\section{Discussion}

In our comparisons  of $Q_{\rm Fe}$ derived from MHD simulations and multispacecraft ICME observations we find several similarities and differences. In Figures~\ref{fig6a} and \ref{fig6b}, the cuts are chosen to approximate the magnetic field rotations seen at each spacecraft.  The ambient level of $Q_{\rm Fe}$ in the data is lower than the BM and higher than the FC model. The specific charge state enhancements  for MC1 are not well matched to the observations.  For example, the STB observations of MC1, which crossed very close to the center of the flux rope, show $Q_{\rm Fe}$ ranging primarily from values of +7 to +12.  The BM results for a similar cut have charge states ranging from +8 to +14 with a distinctive dip at the flux rope center, while the FC results have values that begin at +7 to +8 which then jump to +16  at the location of the flare reconnection heated material.  There are, however, some similarities between MC2 and the FC model, particularly for the ACE and STA cuts.  In the ACE cut, the data show an abrupt jump to charge states of +16 about halfway through the flux rope, which is similar to the jump seen in the FC cut and extends to the end of the flux rope.  At STA the charge state increase occurs about 1/3 of the way through the event and is mostly confined to the center third of the event, in both the data and the FC model results.  

In the global view of the spacecraft data (Figure~\ref{fig3}) we find that the bulk of the charge state enhancements are seen in the back half and surrounding the flux ropes.  For MC1 the enhancements are relatively low and appear in the back half of the flux rope and the region following the flux rope. For MC2, the charge state levels $\langle Q_{\rm Fe} \rangle \ge +12$ are consistent with ICME related material \citep{Lepri2004}. The center of the flux rope contains charge states that are similar to ambient solar wind values, while the enhancements seem to surround the flux rope, being located primarily in the trailing portion of the flux rope, along the western (STA) flank, and in the sheath region ahead of the flux rope structure. 
In the two simulations (Figure~\ref{fig4}) we find that the center of the flux ropes are ``colder" than the ambient solar wind values with charge state enhancements present surrounding and trailing this colder region.  Interestingly, in the BM we see enhancements preceding the flux rope that are very similar to the enhancements seen the sheath region of MC2.  These two flux ropes, particularly MC2, bear an encouraging qualitative resemblance to the model results.

In both events and the two MHD models the charge state enhancements are present most strongly in the second half of the flux rope and in the trailing material.  This result is surprising as previous studies have found that charge state data is more likely to be enhanced near the center of the event, and in particular to be cospatial with the flux rope region\citep{Henke1998,Henke2001,Lynch2003,Richardson2004,Reinard2005,Reinard2008}.   
One possible explanation is that these events illustrate a different or less efficient mechanism that is less common and thus do not stand out in the statistical sampling of \citet{Henke2001} and \citet{Reinard2008}.  Since the associated flares were only small B-class flares and the enhanced charge states only slightly exceed the \citet{Lepri2004} threshold of +12 required for enhanced $\langle Q_{\rm Fe} \rangle$, it is possible that in these events the small associated flares did not have sufficient energy to heat the center of the flux rope. An event associated with a larger M or X-class flare, on the other hand, could experience a more explosive eruption in which the reconnection starts at the center of the event and moves outwards causing a larger flare and a CME with the strongest heating in the center.  In that case, the much larger enhanced charge states in the center of the event would overwhelm the post-eruption flare arcade related heating that we may be seeing here.  

On the other hand, the previous statistical studies may have averaged over the fine structure of the ionic composition enhancements. These studies were based on event averages, which would blur the boundary between enhancements present in an internal ring around the flux rope center and the less enhanced central region, such as that seen in these events (Figure~\ref{fig3}) and the charge states derived from MHD models (Figure~\ref{fig4}). In addition, the previous studies were not designed to examine asymmetries in the structure, so enhancement offsets, such as seen in Figure~\ref{fig3}, would be averaged out.  Additional multispacecraft studies, particularly of CMEs with a large associated flares, would help resolve this issue.  In the meantime, more precise statistical studies are planned, focusing on spatial position within the ICME and relative location of the flare/AR site in order to shed some light on this question.

Recent numerical MHD modeling of sympathetic eruptions by \citet{Toeroek2011} provide a possible scenario to explain the the complex spatial structure of the interplanetary ionic composition enhancements observed in these events. 
Here we refer the reader to Figure~3 of \citet{Toeroek2011} and specifically the location of the current sheets and resulting reconnection regions associated with the sympathetic eruptions of their second and third flux ropes. 
The authors simulate the consecutive, linked eruptions of three flux ropes to mimic the August 1, 2010 series of filament eruptions.  
The consecutive May 19--20, 2007 eruptions resulting in our in-situ observations of the MC1 and MC2 ICMEs bear a striking resemblance to the \citet{Toeroek2011} scenario's second and third filament/flux rope eruptions. 
While the standard eruptive flare (Forbes, 2000 and references therein) current sheet underneath each flux rope eruption supplies the bulk flare heating and enhanced $Q_{\rm Fe}$ signatures to the latter portion of the flux rope ejecta, in the sympathetic eruption scenario, their second eruption current sheet (May 19 CME $\rightarrow$ MC1) also acts as the overlying magnetic breakout reconnection for the third flux rope eruption (May 20 CME $\rightarrow$ MC2). Here the May 19 CME flare-heated $Q_{\rm Fe}$ enhancements occur continuously right up until the eruption of the May 20 CME, providing a natural explanation for the multispacecraft in-situ charge states signatures that seems to link MC1 and MC2 in our Figure~\ref{fig3} spatial mapping results: both a moderate flare-heated $Q_{\rm Fe}$ enhancement in the trailing portion of the flux rope interiors and an extended, continuous enhancement in the solar wind period between the two flux ropes. 
Taking into account the ionic charge state signatures actually allows us to unravel and understand the complex field and plasma signatures in the various spacecraft data sets during our ``dual flux rope" composite/compound event.

It is interesting that the lower half of MC2 (STB measurements) do not show a similar, symmetrical enhancement.  As explained earlier, however, we see that the flare is not symmetrically located under the CME (unlike our simulations). If we assume that the dimming region represents the eruption site, the active region and flare occur to the west of the MC2 source region.  This is consistent with heating of the west side of the CME as seen in in-situ data.  In the MC1 eruption, on the other hand, the active region and flare occur to the east of the source region dimming and it is thus possible that the eastern half of the MC, which is unseen, has more enhancements. 

Finally, we note that ionic and elemental composition can be measured in the corona ``directly" via techniques utilizing UV spectroscopy. The SOHO/UVCS observations, typically combined with other multi-wavelength emission data, can be used to infer the coronal densities and temperatures derived from various line ratios as well as estimating the amount of flare and CME-related heating during the early phases of the CME eruption $\lesssim$2$R_{\odot}$ \citep[see, e.g.,][]{Akmal2001,Ko2003,Bemporad2007,Lee2009,Landi2010}. While these analyses are beyond the scope of the current paper, we acknowledge that bridging the gap between coronal spectroscopy and in-situ composition measurements, both in the solar wind and in CMEs, is an important arena for future research.

\section{Conclusions}

We have described an effort to understand the full global structure of ICMEs from the Sun to the Earth by combining a novel method to derive ionic charge state distributions from MHD models \citep{Lynch2011} with newly developed, groundbreaking multispacecraft data analysis techniques that allow us to construct the global structure of ICMEs \citep{Mulligan2012}. We compare the iron charge states derived from idealized MHD simulations with observations of the May 21-23, 2007 ICMEs.

The large scale comparisons (Figures 3 and 4) show similarities between the $Q_{\rm Fe}$ derived from MHD simulation results and the spatial mapping results derived from STEREO and ACE observations of the May 2007 events. In particular, we find that the charge state enhancements tend to occur at the back end and surrounding the flux rope material. Both the BM and the FC models have similar behavior with enhanced charge states surrounding and following the colder flux rope center. 
Enhancements seen beyond the flux rope in the data, particularly ahead of the MC2 ICME, are well matched to the BM results, suggesting MC1 eruption's eruptive flare reconnection was simultaneously acting as breakout-like reconnection during the sympathetic eruption scenario resulting from the source region's multipolar topology. 

More detailed time series comparisons between the spacecraft observations and cuts through the MHD model results (Figures 6 and 7) show that the ambient solar wind charge state levels are lower than predicted by the BM and higher than predicted by the FC model simulations. 
Quantitative agreement between the observed flux rope iron charge state
distributions are not well captured in the modeling results, particularly
for MC1. However, the FC model does remarkably well at qualitatively
matching the increased charge states observed in MC2, including both the
$Q_{\rm Fe}$ distribution broadening in the ACE data and the bimodal
$Q_{\rm Fe}$ signature in the STA data. Therefore, we find aspects of {\it
both} the BM and FC simulation-derived charges states present in the 2007
May 21--23 multispacecraft ICME observations.

Overall, our analysis and results provide the most comprehensive view to date of the global structure of enhanced composition within ICMEs. More investigation is needed to determine whether the structures observed in this case study are common; in particular, the appearance of enhanced charge states surrounding the flux ropes is at odds with previous statistical results indicating that enhanced charge states are more common at the flux rope center. Finally, we show that the ability to derive charge state information from MHD models provides an important and potentially very useful constraint for these models.

\acknowledgments

A.A.R. and T.M. acknowledge support from NASA SR\&T NNX08AH54G. B.J.L. acknowledges support of AFOSR YIP FA9550-11-1-0048 and NASA HTP NNX11AJ65G. Support for the STEREO mission in-situ data processing and analysis was provided through NASA contracts to the IMPACT (NAS5--03131) and PLASTIC (NAS5--00132) teams. The authors thank the ACE MAG, SWEPAM, and SWICS teams for making their data available on the ACE Science Center Web site\footnote{ \tt{ \url{http://www.srl.caltech.edu/ACE/ASC/} }}.

%
%% The reference list follows the main body and any appendices.
%% Use LaTeX's thebibliography environment to mark up your reference list.
%% Note \begin{thebibliography} is followed by an empty set of
%% curly braces.  If you forget this, LaTeX will generate the error
%% "Perhaps a missing \item?".
%%
%% thebibliography produces citations in the text using \bibitem-\cite
%% cross-referencing. Each reference is preceded by a
%% \bibitem command that defines in curly braces the KEY that corresponds
%% to the KEY in the \cite commands (see the first section above).
%% Make sure that you provide a unique KEY for every \bibitem or else the
%% paper will not LaTeX. The square brackets should contain
%% the citation text that LaTeX will insert in
%% place of the \cite commands.

%% We have used macros to produce journal name abbreviations.
%% AASTeX provides a number of these for the more frequently-cited journals.
%% See the Author Guide for a list of them.

%% Note that the style of the \bibitem labels (in []) is slightly
%% different from previous examples.  The natbib system solves a host
%% of citation expression problems, but it is necessary to clearly
%% delimit the year from the author name used in the citation.
%% See the natbib documentation for more details and options.

\clearpage

%% Use the figure environment and \plotone or \plottwo to include
%% figures and captions in your electronic submission.
%% To embed the sample graphics in
%% the file, uncomment the \plotone, \plottwo, and
%% \includegraphics commands
%%
%% If you need a layout that cannot be achieved with \plotone or
%% \plottwo, you can invoke the graphicx package directly with the
%% \includegraphics command or use \plotfiddle. For more information,
%% please see the tutorial on "Using Electronic Art with AASTeX" in the
%% documentation section at the AASTeX Web site,
%% http://www.journals.uchicago.edu/AAS/AASTeX.
%%
%% The examples below also include sample markup for submission of
%% supplemental electronic materials. As always, be sure to check
%% the instructions to authors for the journal you are submitting to
%% for specific submissions guidelines as they vary from
%% journal to journal.

%% This example uses \plotone to include an EPS file scaled to
%% 80% of its natural size with \epsscale. Its caption
%% has been written to indicate that additional figure parts will be
%% available in the electronic journal.

\begin{figure}
\begin{center}
%$
%\begin{array}{cc}
%\includegraphics[width=25pc]{f1a.eps} & \includegraphics[width=25pc]{f1b.eps}
%\end{array}
%$
\includegraphics[width=20pc]{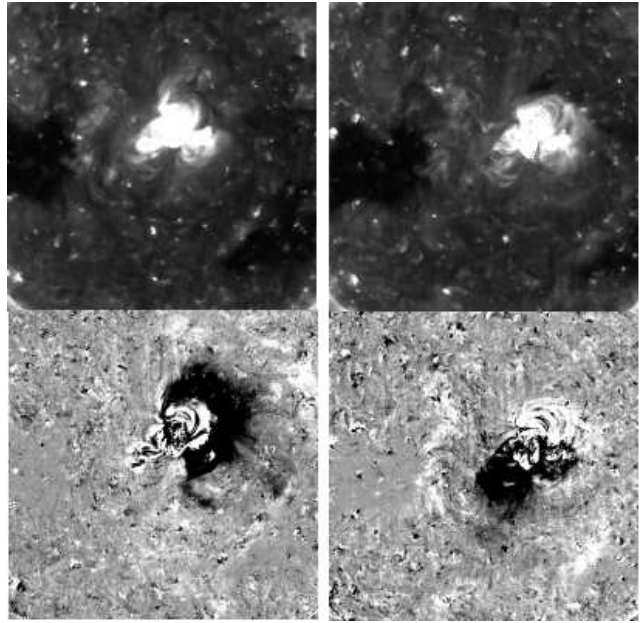}

\end{center}
\caption{EUVI 171\AA\ base (top) and difference (bottom) images showing changes in the source region following CME eruptions on May 19 (left) and May 20 (right). These dimmings indicate where the majority of the CME plasma originated. In the case of the May 19 (MC1) event, the eruption took place primarily to the northwest of the active region, while in the May 20 (MC2) event the eruption was located to the southeast of the active region.  \label{fig1}
}
\end{figure}

\begin{figure}
\includegraphics[width=39pc]{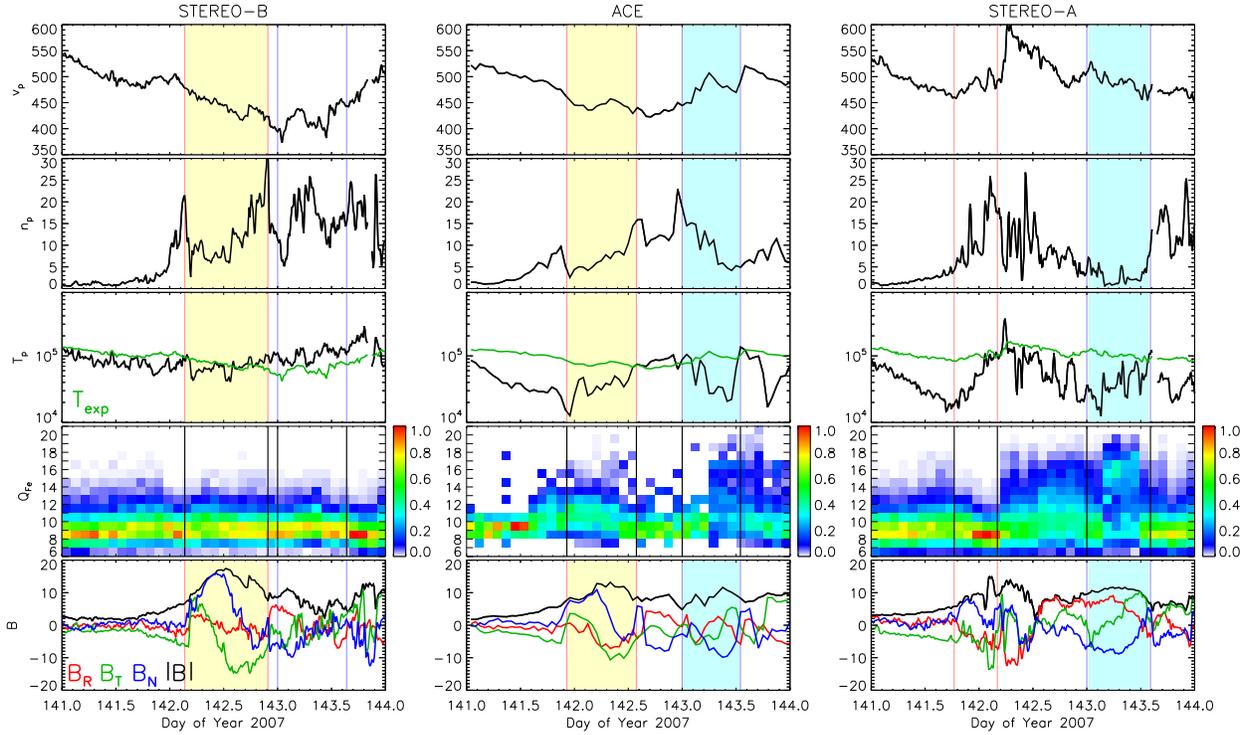}
\caption{STEREO-B (left), ACE (middle), and STEREO-A (right) spacecraft data for 2007 May 21--24 (DOY 141--144). From top to bottom shows $v_p$, $n_p$, $T_p$, and the empirical ``expected" temperature $T_{\rm exp}$ (green), distribution of iron charge states (from +6 to +20), and the RTN magnetic field components $B_{\rm R}$ (red), $B_{\rm T}$ (green), $B_{\rm N}$ (blue), and $B$ magnitude (black). The event interval MC1 (MC2) is indicated as vertical yellow (light blue) lines with trajectories that intersect the ICME flux rope shown as the shaded yellow (light blue) bars.  \label{fig2}
}
\end{figure}

\begin{figure}
\centerline{\includegraphics[width=45pc]{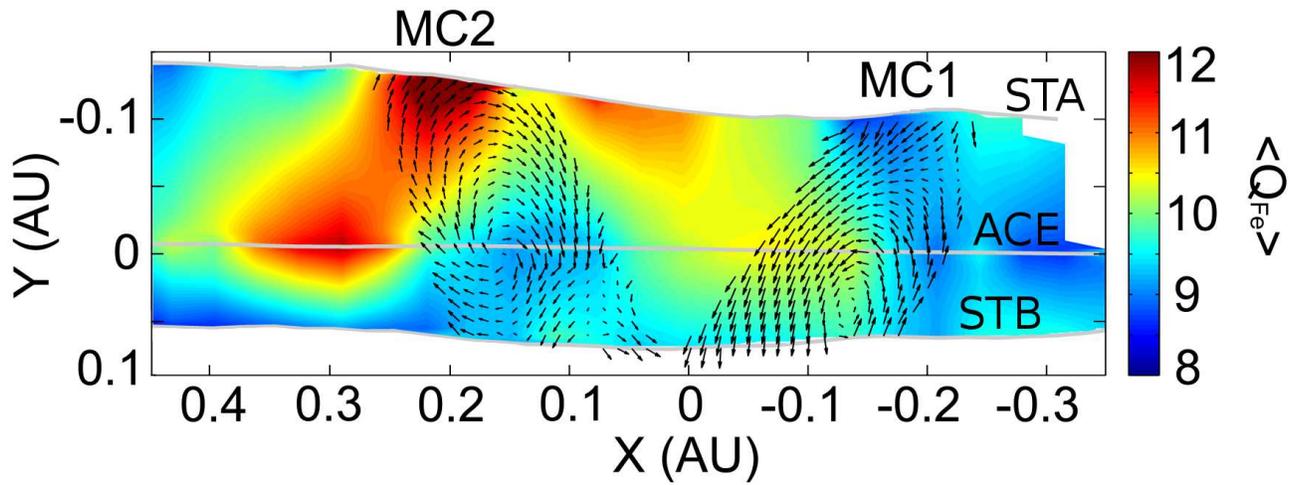}}
\caption{Application of the \citet{Mulligan2012} spatial mapping interpolation technique to the average iron charge state $\langle Q_{\rm Fe} \rangle$ data for the 2007 May 21--24 interval. The gray lines show, from top to bottom, the STEREO-A, ACE,  and STEREO-B spacecraft tracks, i.e. the position of the each spacecraft's measured data.  The vector field arrows show the magnetic field in the ecliptic plane during the MC1 and MC2 regions indicated in Figure~\ref{fig2}. \label{fig3}
}
\end{figure}

\begin{figure}
\includegraphics[width=39pc]{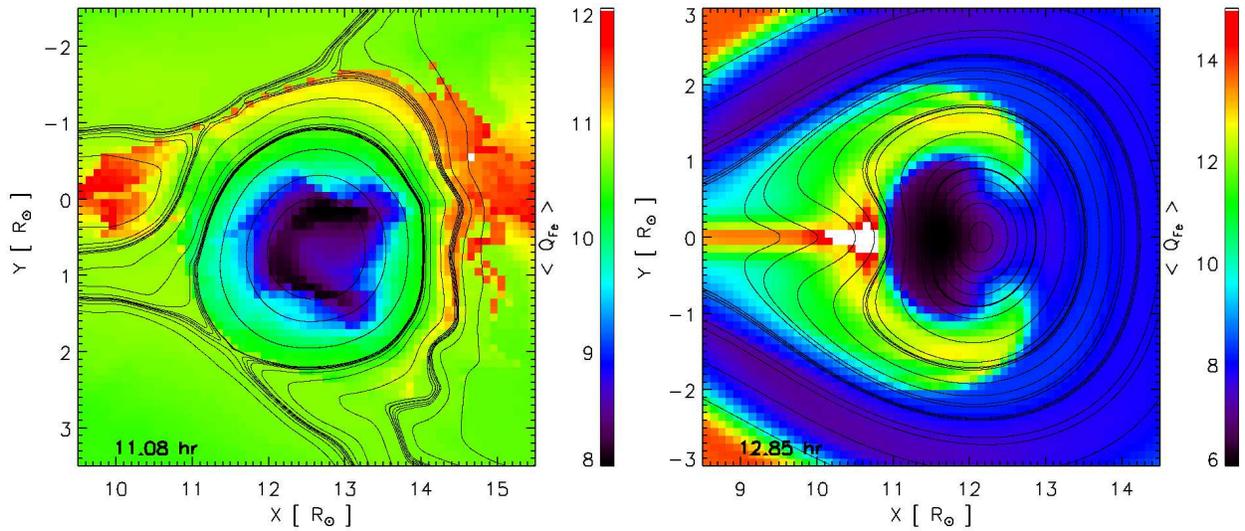}
\caption{Snapshot of the average iron charge state $\langle Q_{\rm Fe} \rangle$ spatial distribution derived from 2D MHD simulations of CME initiation: Left panel, the breakout model (BM); Right panel, the flux cancellation (FC) model. The simulation flux ropes are at $\sim$12$R_{\odot}$ in their respective domains but the heavy ion charge states have already frozen-in. Snapshots were selected so that each flux rope was at or past 12 $R_{\odot}$.  See \citet{Lynch2011} for further details. \label{fig4}
}
\end{figure}

\begin{figure}
\includegraphics[width=32pc]{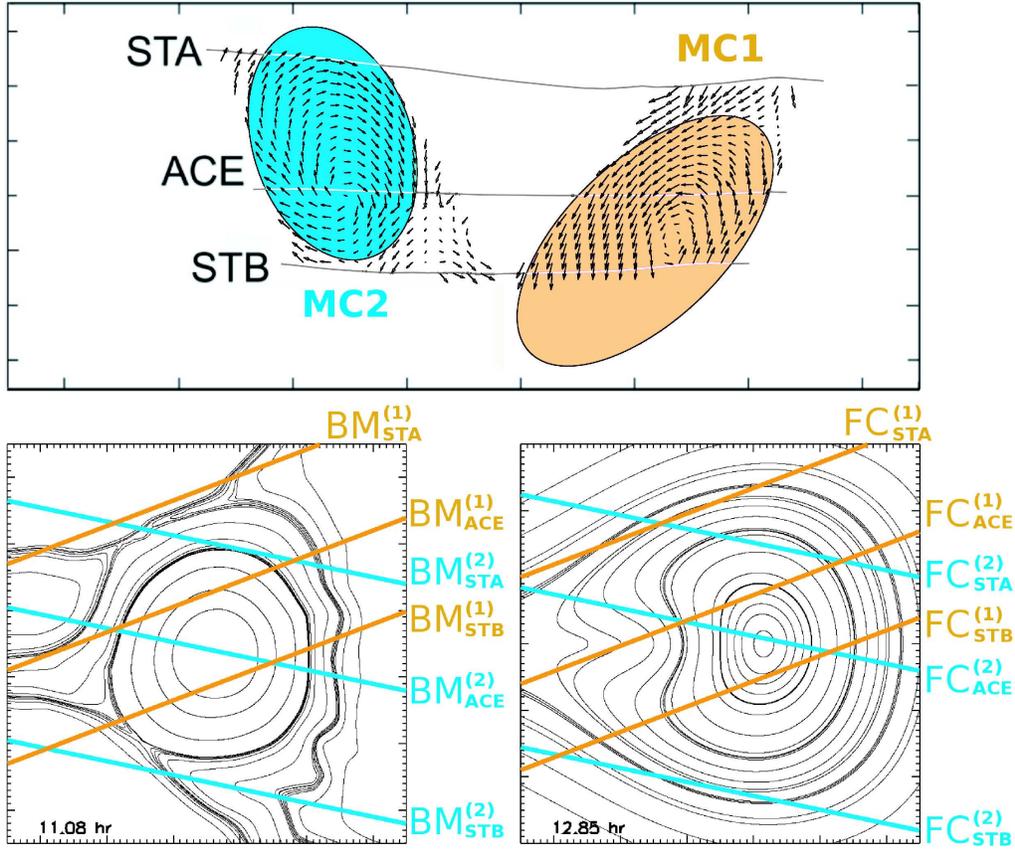}
\caption{Top panel plots a portion of the multispacecraft spatial mapping tracks and the ecliptic plane field rotations for the two ICME intervals \citep[adapted from][]{Mulligan2012}. Lower left panel shows the synthetic spacecraft trajectory cuts through the breakout model simulation corresponding to MC1 (${\rm BM}^{(1)}$) and MC2 (${\rm BM}^{(2)}$). Lower right panel shows the trajectories through the flux cancellation simulation (${\rm FC}^{(1)}$, ${\rm FC}^{(2)}$). \label{fig5}
}
\end{figure}

\begin{figure}
\includegraphics[width=22pc]{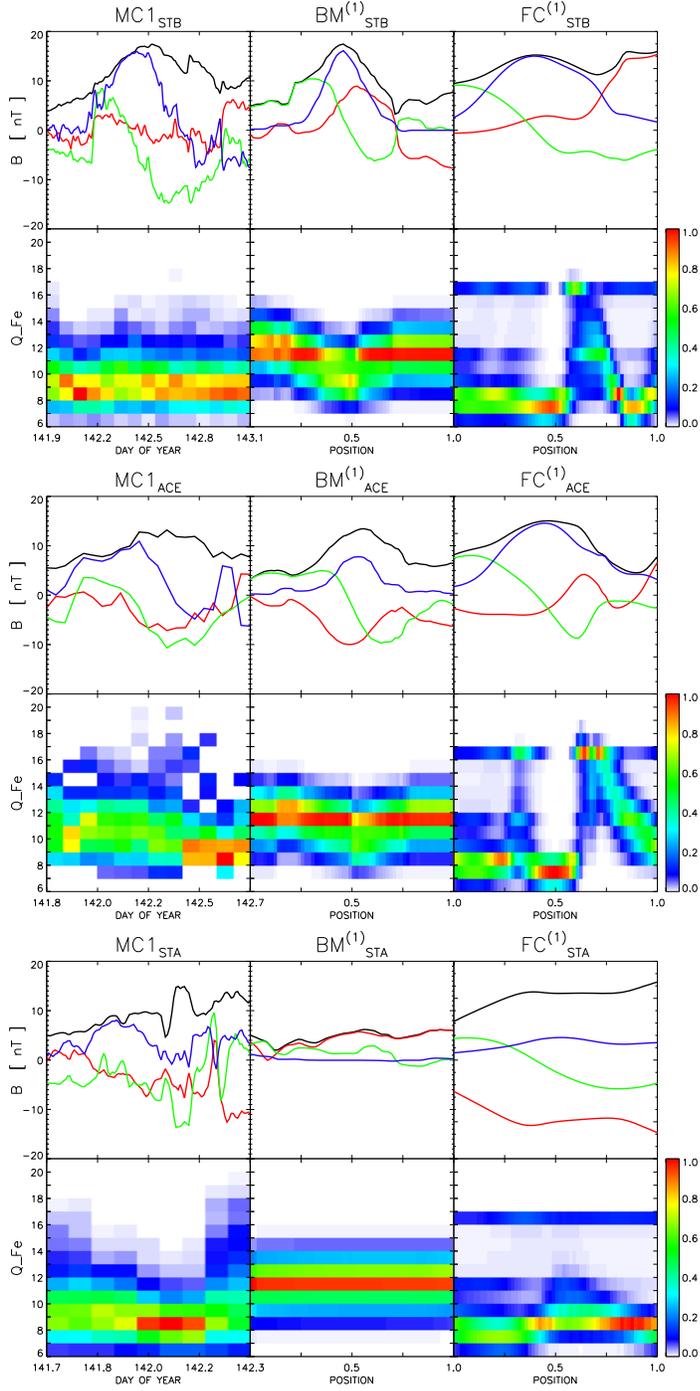}
\caption{Comparison between the MC1 multispacecraft data and the MC1 simulation trajectory samples shown in Figure~\ref{fig5}. The top, middle, bottom plots show the intervals for STEREO-B, ACE, and STEREO-A respectively. Each plot shows the observed magnetic field (upper panel), the extended $Q_{\rm Fe}$ distribution (lower panel), and the corresponding quantities in the breakout model (BM) and flux cancellation (FC) simulations. In order to facilitate the comparison between the observations and simulation data we have transformed the simulation magnetic field profiles to match the sign of the axial flux rope field and sense of rotation; see text for details. \label{fig6a}
}
\end{figure}

\begin{figure}
\includegraphics[width=22pc]{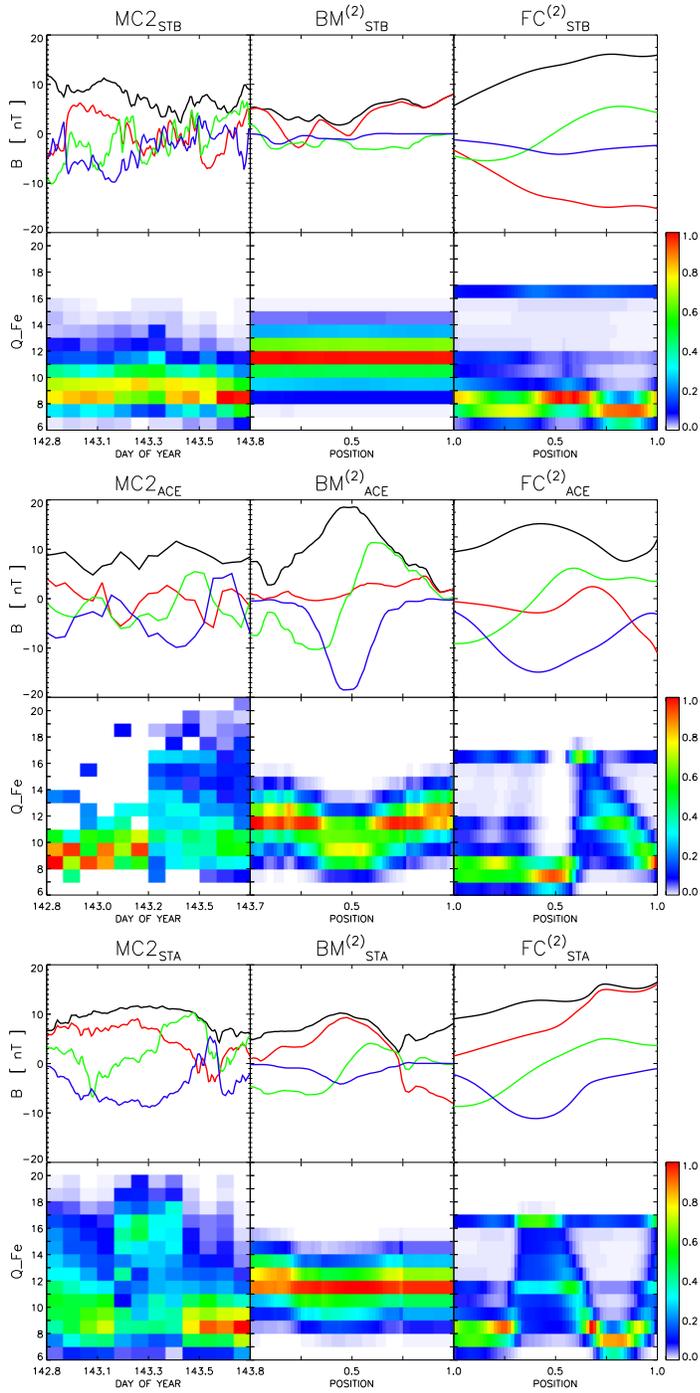}
\caption{Comparison between the MC2 multispacecraft data and the MC2 simulation trajectory samples shown in Figure~\ref{fig5} in the same format as Figure~\ref{fig6a}. \label{fig6b}
}
\end{figure}

%%\begin{figure}
%\includegraphics[angle=90,scale=.50]{f3.eps}
%\caption{Animation still frame taken from \citet{kim03}.
%This figure is also available as an mpeg
%animation in the electronic edition of the
%{\it Astrophysical Journal}.}
%\end{figure}

%% If you are not including electonic art with your submission, you may
%% mark up your captions using the \figcaption command. See the
%% User Guide for details.
%%
%% No more than seven \figcaption commands are allowed per page,
%% so if you have more than seven captions, insert a \clearpage
%% after every seventh one.

%% The following command ends your manuscript. LaTeX will ignore any text
%% that appears after it.

\end{document}